\documentclass{moriond}

\usepackage{amssymb}

\def\be{\begin{equation}}
\def\ee{\end{equation}}
\def\bea{\begin{eqnarray}}
\def\eea{\end{eqnarray}}

\newcommand{\alphas}{\alpha_{\rm s}}
\newcommand{\alphasmZ}{\alphas(\rm m^2_{_{\rm Z}})}
\newcommand{\sqrts}{\sqrt{\rm s}}

\newcommand{\lqcd}{\Lambda_{_{\rm QCD}}}

\newcommand{\epem}{e^+e^-}
\newcommand{\meff}{m{_{\rm eff}}}

\providecommand{\bbbar}{\rm b\overline{b}}

\newcommand{\Photo}{\includegraphics[height=30mm]{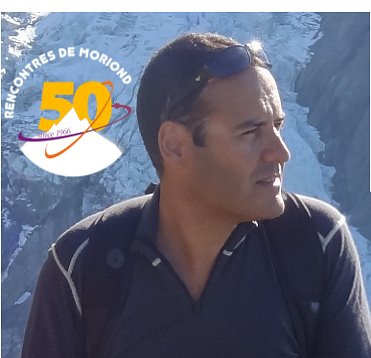}}

\begin{document}
\title{$\alphas$ determination at NNLO$^\star$+NNLL accuracy from the\\ energy evolution of jet fragmentation functions at low $z$}


\author{\underline{David d'Enterria}$^1$, and Redamy P\'erez-Ramos$^{2,3}$.}

\address{$^1$  CERN, PH Department, CH-1211 Geneva 23, Switzerland\\
$^2$ Sorbonne Universit\'e, UPMC Univ Paris 06, UMR 7589, LPTHE, F-75005, Paris, France\\
CNRS, UMR 7589, LPTHE, BP 126, 4 place Jussieu, F-75252 Paris Cedex 05, France}

\maketitle\abstracts{
The QCD coupling $\alphas$ is extracted from the energy evolution of the first two moments (multiplicity and
mean) of the parton-to-hadron fragmentation functions at low fractional hadron momentum $z$.
Comparisons of the experimental $\epem$ and deep-inelastic $e^\pm$p jet data to our NNLO$^*$+NNLL predictions, 
allow us to obtain $\alphasmZ$~=~0.1205$\pm$0.0010$^{+0.0022}_{-0.0000}$, in excellent 
agreement with the current world average determined using other methods at the same level of accuracy.}

\section{Introduction}

For massless quarks and fixed number of colours, the only fundamental parameter of the theory of the strong
interaction, quantum chromodynamics (QCD), is its coupling constant $\alphas$. 
Starting from a value of $\lqcd\approx$~0.2~GeV, where the perturbatively-defined coupling diverges, 
$\alphas$ decreases with increasing energy $Q$ following a $1/\ln(Q^2/\lqcd^2)$ dependence. The current
uncertainty on $\alphas$ evaluated at the Z mass, $\alphasmZ$~=~0.1185$\pm$0.0006, is $\pm$0.5\%~\cite{PDG}, making
of the strong coupling the least precisely known of all fundamental constants in nature. Improving our knowledge
of $\alphas$ is a prerequisite to reduce the uncertainties in perturbative-QCD calculations 
of all (partonic) cross sections at hadron colliders, and for precision fits of the Standard Model
($\alphas$ dominates e.g. the Higgs boson H$\to\bbbar$ partial width uncertainty). It has also 
far-reaching implications including the stability of the electroweak vacuum~\cite{Buttazzo:2013uya} or 
the scale at which the interaction couplings unify.\\

Having at hand new independent approaches to determine $\alphas$ from the data, with experimental and
theoretical uncertainties different from those of the other methods currently used, is crucial to reduce the
overall uncertainty in the combined $\alphas$ world-average value~\cite{PDG}. 
In Refs.~\cite{NMLLA_NLO,NMLLA_NLO2} we have presented a novel technique to extract $\alphas$ from the energy
evolution of the moments of the parton-to-hadron fragmentation functions (FF) 
computed at approximate next-to-leading-order (NLO$^\star$) accuracy including next-to-next-to-leading-log (NNLL)
resummation corrections. This approach is extended here to include full NLO, plus a set of NNLO, corrections.
Our new NNLO$^\star$+NNLL theoretical results for the energy dependence of the hadron multiplicity and mean value
of the FF are compared to jet fragmentation measurements in $\epem$ and deep-inelastic $e^\pm$p collisions, in
order to extract a high-precision value of $\alphas$.

\section{Evolution of the parton-to-hadron fragmentation functions at NNLO$^\star$+NNLL}

The distribution of hadrons in a jet is encoded in a fragmentation function, 
$D_{\rm i\to h}(z,Q)$, which describes the probability that the parton $i$ fragments
into a hadron $h$ carrying a fraction $z=p_{\rm hadron}/p_{\rm parton}$ of the parent parton's momentum.
Usually one writes the FF as a function of the log of the inverse of $z$, $\xi=\ln(1/z)$, to 
emphasize the region of relatively low momenta that dominates the jet hadronic fragments. 
Starting with a parton at a given 
energy $Q$, its evolution to another energy scale $Q'$ is driven by a branching process of parton radiation
and splitting, resulting in a jet ``shower'', which can be computed 
perturbatively using the DGLAP~\cite{dglap} equations at large $z\gtrsim 0.1$, and the Modified
Leading Logarithmic Approximation (MLLA)~\cite{mlla}, resumming soft and collinear singularities at small
$z$. Due to colour coherence and gluon radiation interference, not the softest partons but those with
intermediate energies multiply most effectively in QCD cascades, leading to a final FF with a typical
``hump-backed plateau'' (HBP) shape as a function of $\xi$ (Fig.~\ref{fig:1}), which can be expressed in terms
of a distorted Gaussian:  
\begin{equation}
D(\xi,Y,\lambda) = {\cal N}/(\sigma\sqrt{2\pi})\cdot e^{\left[\frac18k-\frac12s\delta-
\frac14(2+k)\delta^2+\frac16s\delta^3+\frac1{24}k\delta^4\right]}\,, \mbox{ with }\delta=(\xi-\bar\xi)/\sigma,
\label{eq:DG}
\end{equation}
where ${\cal N}$ is the 
average hadron multiplicity inside a jet, and $\bar\xi$, $\sigma$, $s$, and $k$ are respectively the mean peak
position, dispersion, skewness, and kurtosis of the distribution.
The set of integro-differential equations for the FF evolution combining hard (DGLAP, MLLA, next-to-MLLA) and soft (DLA) 
radiation can be solved by expressing the Mellin-transformed hadron distribution in terms of the anomalous
dimension $\gamma$: $D\simeq C(\alphas(t))\exp\left[\int^t \gamma(\alphas(t')) dt\right]$ for $t=\ln Q$, 
leading to a perturbative expansion in half powers of $\alphas$: 
$\gamma\sim {\cal O}(\alphas^{^{1/2}})+{\cal O}(\alphas)+{\cal O}(\alphas^{^{3/2}})+{\cal O}(\alphas^{^2})+{\cal O}(\alphas^{^{5/2}})+\cdots$.
Corrections up to order $\alphas^{^{3/2}}$ were computed for the first time in Ref.~\cite{NMLLA_NLO,NMLLA_NLO2}.
The full set of NLO ${\cal O}(\alphas^2)$ terms for the anomalous dimension, including the two-loop splitting
functions, plus a fraction of the ${\cal O}(\alphas^{^{5/2}})$ terms, coming from the NNLO expression for the
$\alphas$ running, have now been computed~\cite{DdERPR}. Upon inverse-Mellin transformation, one obtains
the energy evolution of the FF, and its associated moments, 
as a function of Y~=~$\ln(E/\lqcd)$, for an initial parton energy $E$, down to a shower cut-off
scale $\lambda = \ln(Q_{_{0}}/\lqcd)$ for $N_f=3,4,5$ quark flavors.
The resulting formulae for the energy evolution of the moments depend on $\lqcd$ as {\it single} free
parameter. Particularly simple expressions are obtained in the limiting-spectrum case ($\lambda=0$,
i.e. evolving the FF down to Q$_0 = \lqcd$) motivated by the ``local parton hadron duality'' hypothesis for infrared-safe observables 
which states that the distribution of partons in jets are simply renormalized in the hadronization process
without changing their HBP shape. Thus, by fitting  the experimental hadron jet data at various energies to
Eq.~(\ref{eq:DG}), one can determine $\alphas$ from the corresponding energy-dependence of its FF 
moments.

\section{Data-theory comparison and $\alphas$ extraction}

\begin{figure}[htpb!]
\centerline{
\includegraphics[width=0.50\linewidth,natwidth=787,natheight=465]{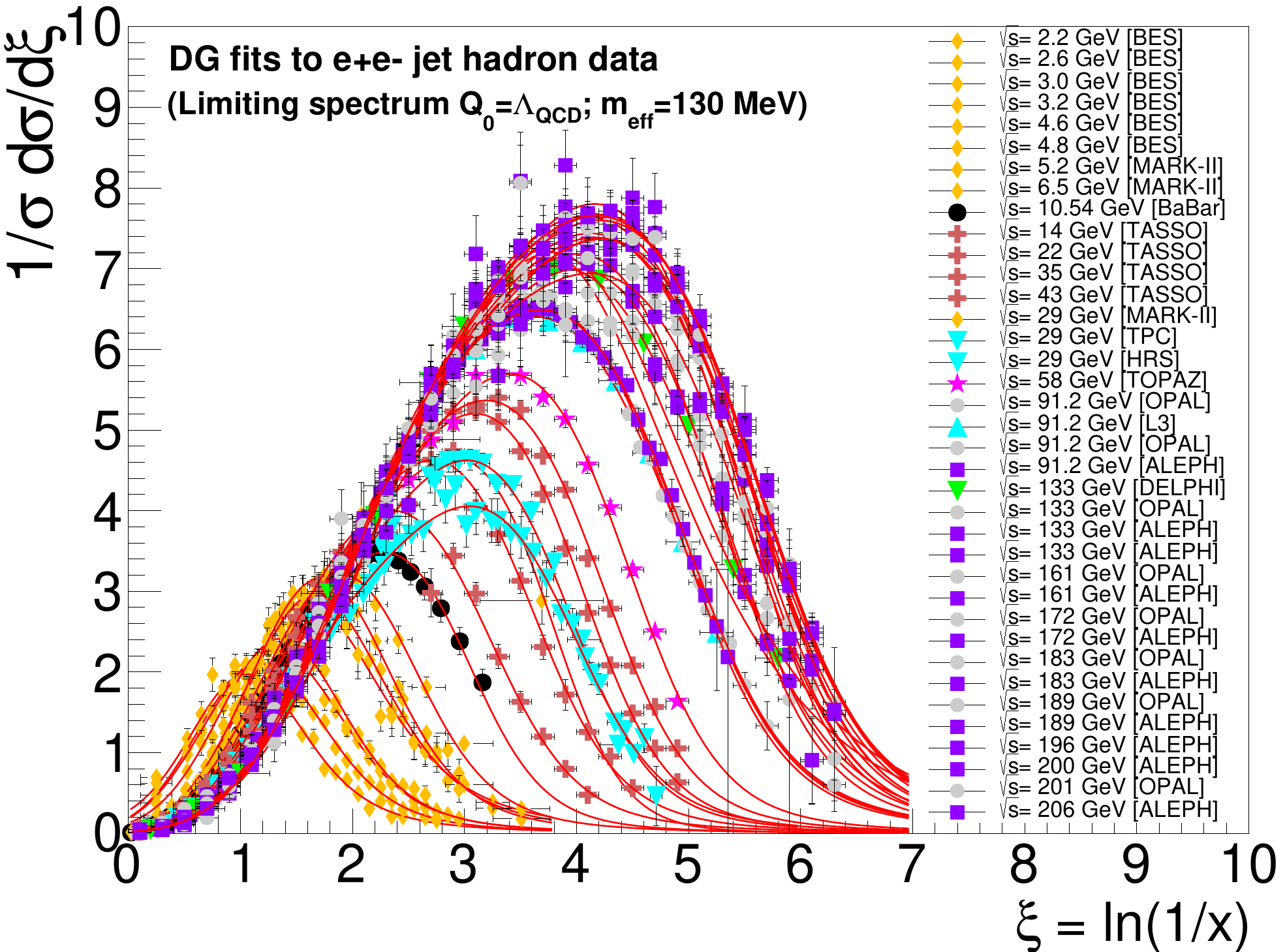}
\includegraphics[width=0.50\linewidth,natwidth=787,natheight=465]{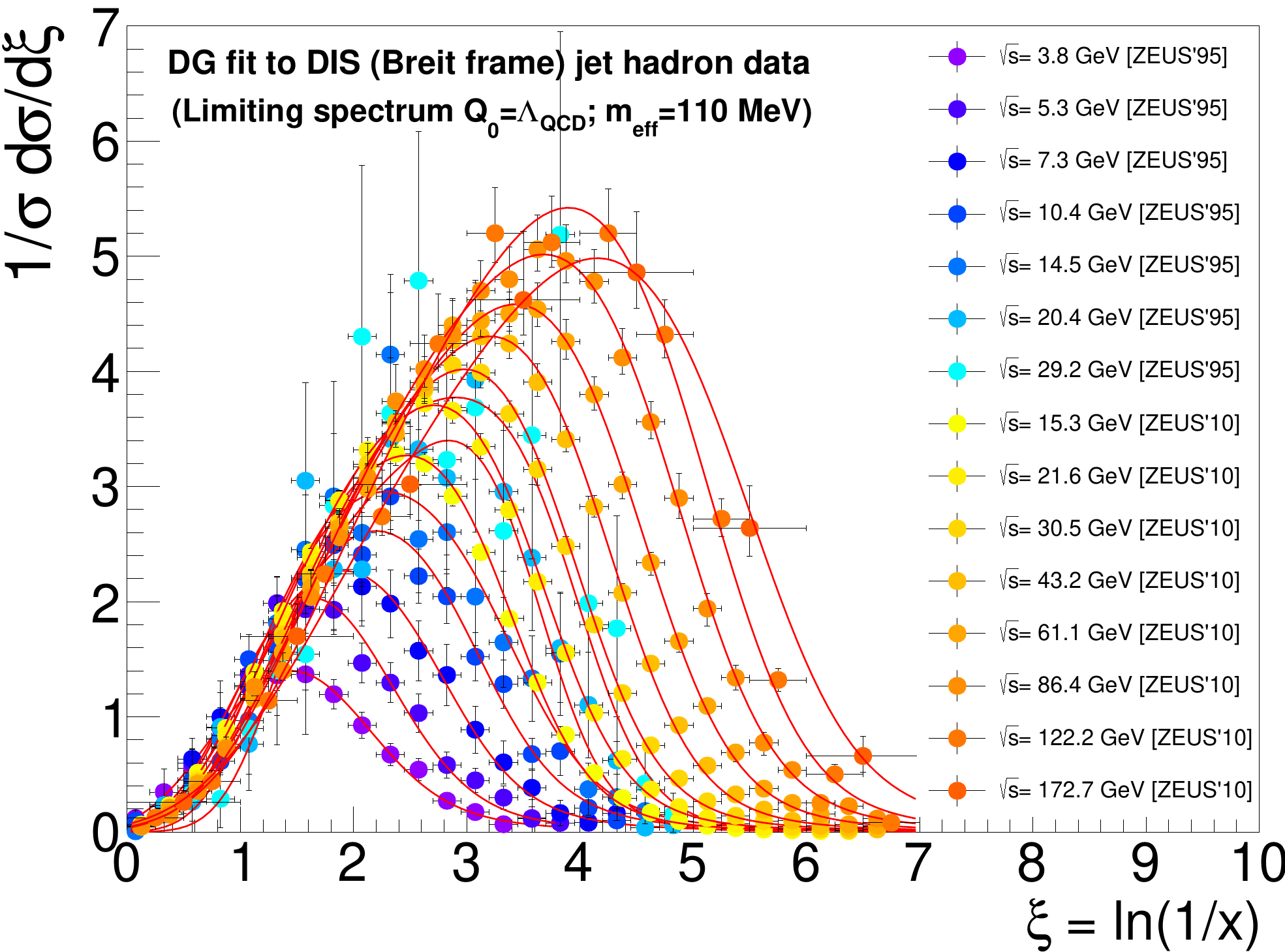}
}
\caption[]{Charged-hadron distributions in jets as a function of $\xi=\ln(1/z)$ measured 
in $\epem$ at $\sqrts\approx$~2--200~GeV (left) and $e^\pm,\nu$-p (Breit frame, scaled up by $\times$2 to
account for the full hemisphere) at $\sqrts\approx$~4--180~GeV (right), individually fitted to
Eq.~(\ref{eq:DG}), with the hadron mass corrections ($\meff$~=~130,110~MeV) quoted.}
\label{fig:1}
\end{figure}

The first step of our procedure is to fit to Eq.~(\ref{eq:DG}) all existing jet FF data measured in $\epem$
and $e^\pm,\nu$-p collisions at $\sqrts\approx$~1--200~GeV (Fig.~\ref{fig:1}) in order to obtain the
corresponding FF moments at each jet energy.
Finite hadron-mass effects in the DG fit are accounted for through a rescaling of the theoretical (massless)
parton momenta with an effective mass $\meff$ as discussed in Refs.~\cite{NMLLA_NLO,NMLLA_NLO2}. The overall
normalization of the HBP spectrum (${\cal K}_{\rm ch}$), which determines the average charged-hadron
multiplicity of the jet, is an extra free parameter in the DG fit which, nonetheless, plays no role in the
final $\lqcd$ value given that its extraction {\em just} depends on the {\it evolution} of the multiplicity, 
and not on its absolute value at any given energy. 
\begin{figure}[htpb!]
\includegraphics[width=0.50\linewidth,height=5.8cm]{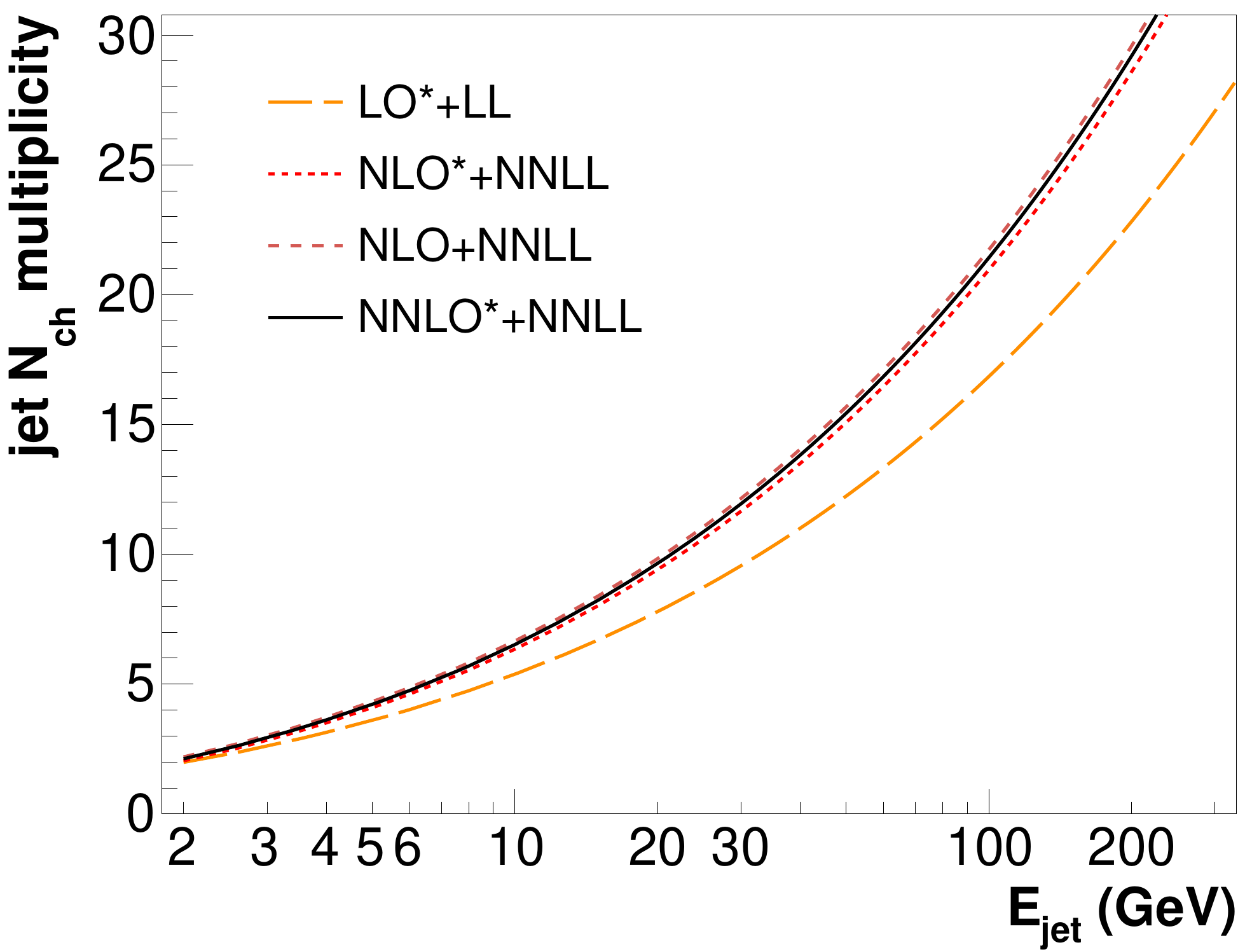}
\includegraphics[width=0.50\linewidth,height=5.8cm]{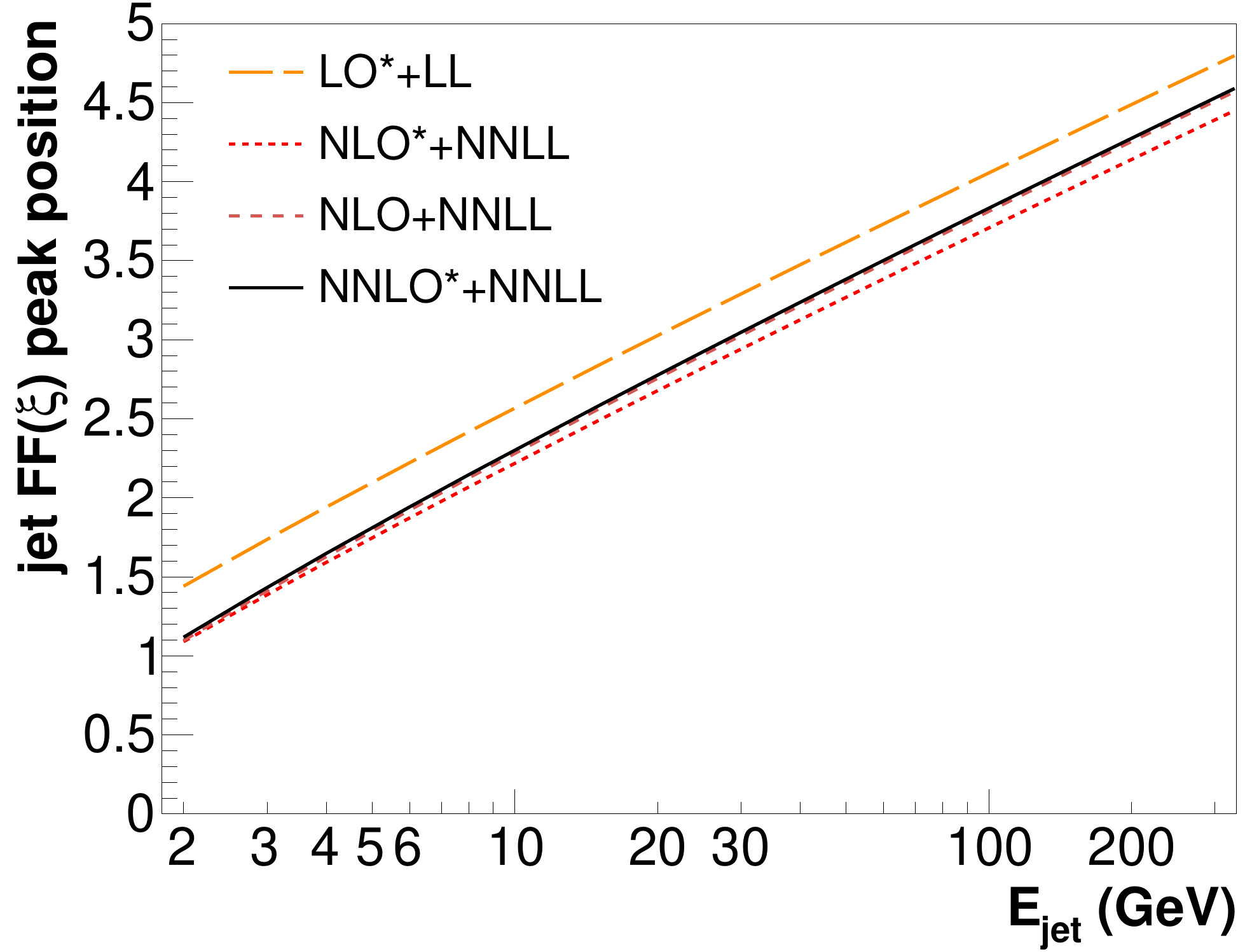}
\includegraphics[width=0.50\linewidth]{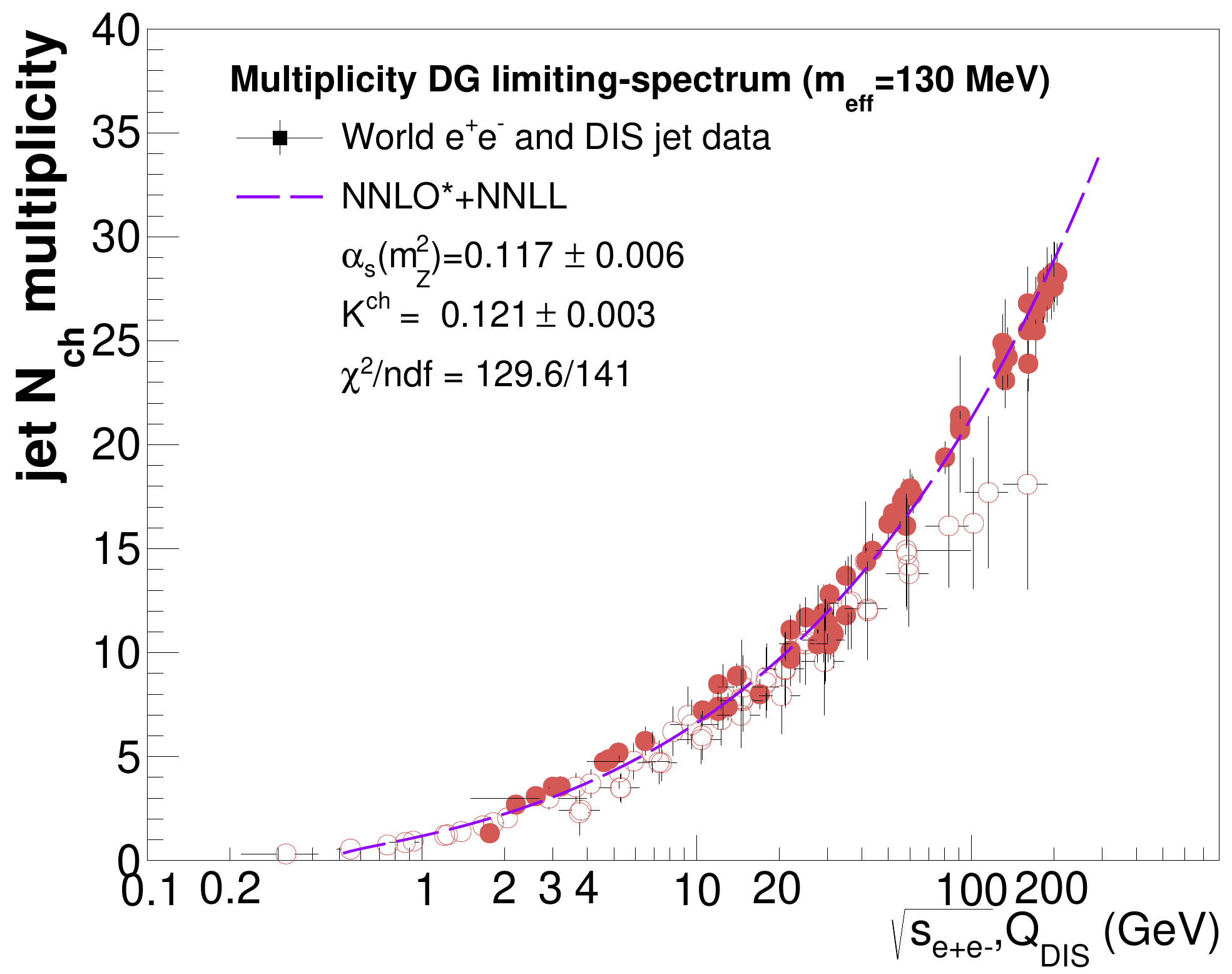}
\includegraphics[width=0.50\linewidth]{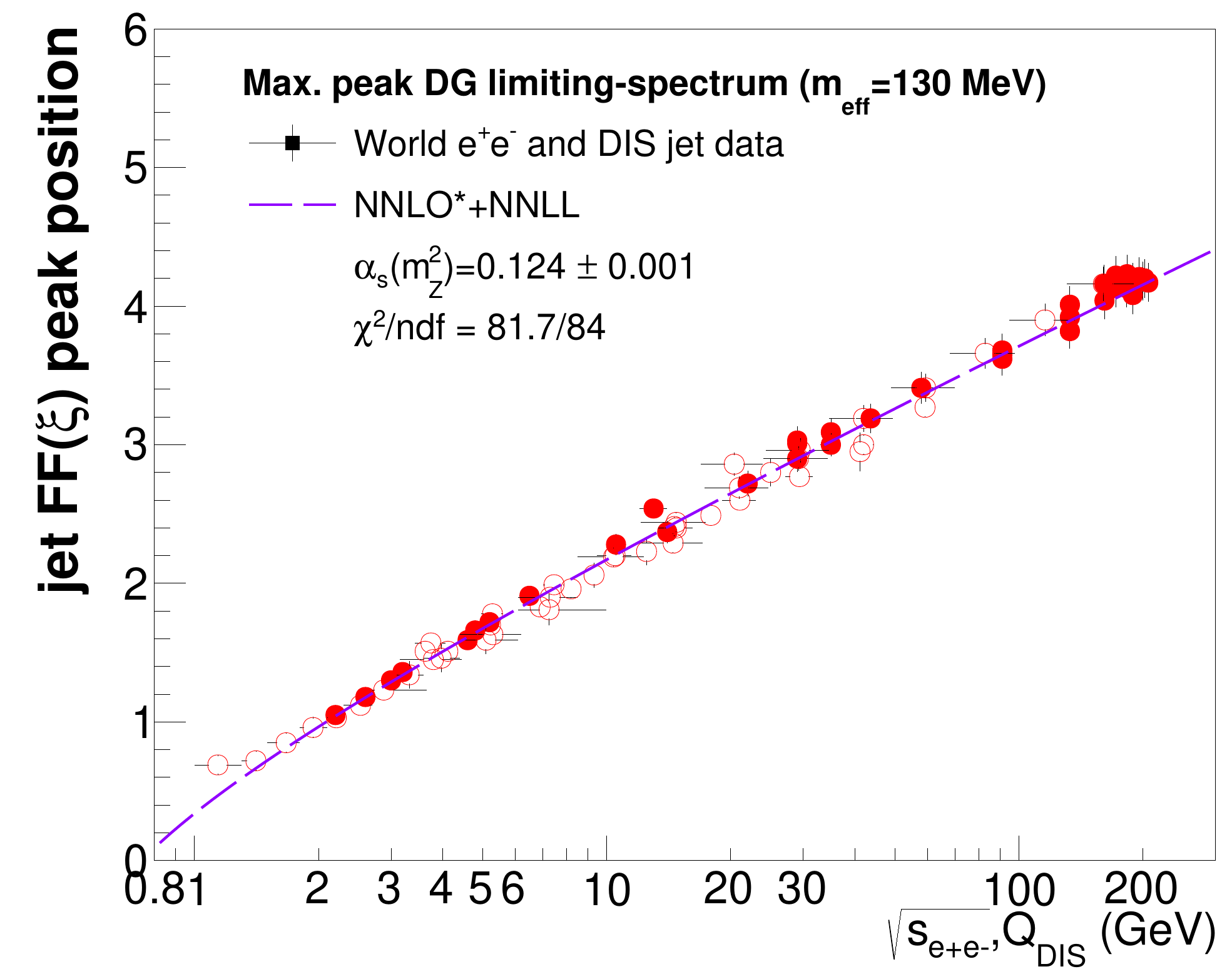}
\caption[]{Energy evolution of the jet charged-hadron multiplicity (left) and FF peak position (right).
Top: Comparison of theoretical predictions at four levels of accuracy.
Bottom: Fit of the experimental $\epem$ and DIS jet data to the NNLO$^\star$+NNLL predictions.
The obtained ${\cal K}_{\rm ch}$ normalization constant, 
the individual NNLO$^\star$ $\alphasmZ$ values, and the goodness-of-fit per degree-of-freedom $\chi^2/$ndf, 
are quoted.}
\label{fig:2}
\end{figure}

\begin{figure}[htpb!]
\centerline{
\includegraphics[width=0.65\linewidth]{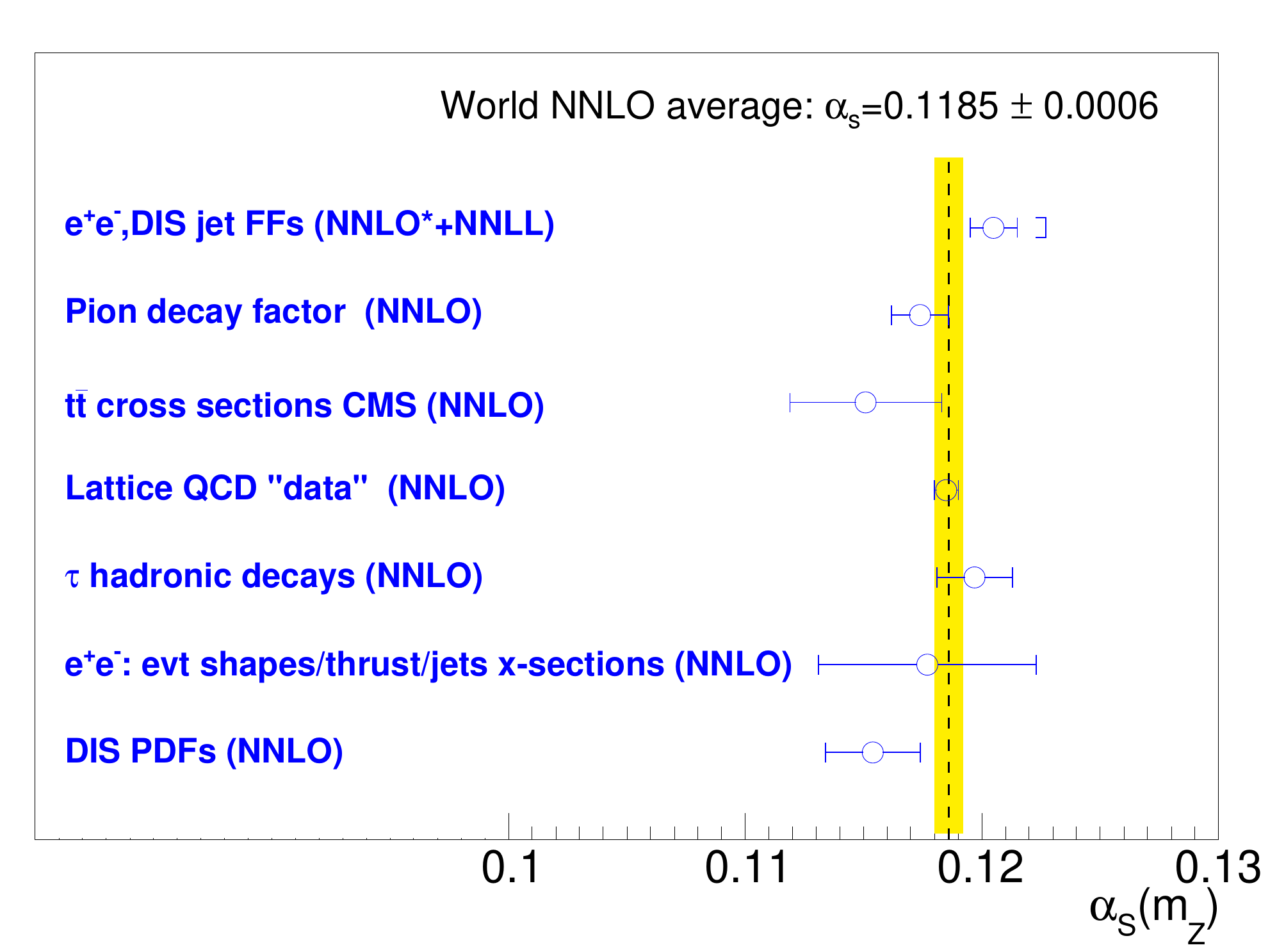}
}
\caption[]{Summary of $\alphas$ determinations using different methods at NNLO$^{(\star)}$ accuracy.
The dashed line and shaded (yellow) band indicate the current world-average and uncertainty (listed also on the top)~\cite{PDG}.
}
\label{fig:alphas_NLO}
\end{figure}

Once the FF moments have been obtained, one can perform a combined fit of them as a function of the original
parton energy (which in the case of $\epem$ corresponds to half the centre-of-mass energy $\sqrts/2$ and, for
DIS, to the invariant four-momentum transfer $Q_{_{\rm DIS}}$). The top panels of Fig.~\ref{fig:2} show the energy
evolution of the zeroth (multiplicity) and first (peak position, closely connected to the mean of the
distribution) moments of the FF, computed at an increasingly higher level of accuracy (from LO 
to NNLO$^\star$). 
The hadron multiplicity and FF peak increase exponentially and logarithmically with energy, and the theoretical
convergence of their evolutions are very robust as indicated by the small changes introduced by incorporating 
higher-order terms. The corresponding data-theory comparisons are seen in the 
bottom panels of Fig.~\ref{fig:2}. The NNLO$^\star$+NNLL limiting-spectrum ($\lambda$~=~0) predictions
for $N_f$~=~5 active quark flavours\footnote{The moments of the lowest-$\sqrts$ data have
a few-percent correction applied to account for the (slightly) different ($N_f$~=~3,4) evolutions below the charm and
bottom production thresholds.}, leaving $\lqcd$ as a free parameter, reproduce very well the data.
The most ``robust'' FF moment for the determination of $\lqcd$ is the peak position $\xi_{\rm max}$ which is
quite insensitive to most of the uncertainties associated with the extraction method (DG and energy evolution
fits, finite-mass corrections)~\cite{NMLLA_NLO2} as well as to higher-order corrections. The hadron multiplicities
measured in DIS jets appear somewhat smaller (especially at high energy) than those measured in $\epem$ collisions, 
due to limitations in the FF measurement only in half (current Breit) $e^\pm$p hemisphere and/or in the
determination of the relevant $Q$ scale~\cite{NMLLA_NLO2}. 
The value of $\alphasmZ$ obtained from the combined multiplicity+peak fit yields
$\alphasmZ$~=~0.1205$\pm$0.0010, where the error includes all sources of uncertainty discussed
in Ref.~\cite{NMLLA_NLO2}. 
An extra theoretical scale uncertainty of $^{+0.0022}_{-0.0000}$ should be added
(this is a conservative estimate obtained only at NLO$^\star$ accuracy~\cite{NMLLA_NLO2}).
In Fig.~\ref{fig:alphas_NLO} we compare our extracted $\alphasmZ$ value to all other NNLO results
from the latest PDG compilation~\cite{PDG}, plus that obtained from the pion decay factor~\cite{Kneur:2013coa}, and the
top-quark pair production cross sections~\cite{Chatrchyan:2013haa}. The precision of our result ($^{+2\%}_{-1\%}$) is clearly
competitive with the other measurements, with a totally different set of experimental and theoretical
uncertainties. A simple weighted average of all these NNLO values yields: $\alphasmZ$~=~0.1186$\;\pm\;$0.0004,
in perfect agreement with the central value of the current world-average, $\alphasmZ$~=~0.1185$\pm$0.0006, but
with a 30\% smaller uncertainty.
\vspace{-0.4cm}

We are grateful to Anatoly~Kotikov and Gavin~Salam for useful discussions.
R.~P\'erez-Ramos thanks the support from the CERN TH Division where part of this work was carried out.

\end{document}